\documentclass[sigconf,nonacm]{acmart}

\AtBeginDocument{%
 }

\usepackage{xurl}

\begin{document}

\title{On the Meaning of the Web as an Object of Study}

\author{Claudio Gutierrez}
\orcid{0000-0002-4559-6544}
\affiliation{%
  \institution{Universidad de Chile}
  \department{Department of Computer Science \& IMFD}
  \city{Santiago}
  \country{Chile}
}

\author{Daniel Hern\'{a}ndez}
\orcid{0000-0002-7896-0875}
\affiliation{%
  \institution{University of Stuttgart}
  \department{Institute for Artificial Intelligence, Analytic Computing}
  \city{Stuttgart}
  \country{Germany}
}


\begin{abstract}
  This text advances the hypothesis that the meaning of the Web as an object of study has diluted as a clear research domain. One example of this phenomenon is the identity crisis of the Web Conference and the International Semantic Web Conference. At its root is the Web's evolution from a focused technological object into a universal digital environment, a transition whose very success has fragmented its academic community and obscured its core identity. We chart this trajectory from a well-defined object of study to a fragmented backdrop, identifying key pressures such as the ``academic tragedy of the commons'' and the disruptive force of AI. We conclude that a fundamental community discussion is needed to define what it means to study the Web now that it has become the universal infrastructure for global digital activity.
\end{abstract}

%

\keywords{Web Conferences, Digital Environment, Disciplinary Dilution}


\maketitle

\noindent
\textbf{Preprint.} The published version appears in the poster track of the 18th ACM Web Science Conference Companion (WebSci Companion '26). DOI: \url{https://doi.org/10.1145/3795513.3807425}

\smallskip
\noindent
\includegraphics[height=1em]{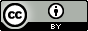} This work is licensed under a \href{https://creativecommons.org/licenses/by/4.0/}{Creative Commons Attribution 4.0 International License}.

\section*{Introduction}

The World Wide Web, conceived as a universal space of interconnected information \cite{bernersLee1999}, has long surpassed its status as a technological novelty (in its origins in the 1990s) to become an omnipresent and nearly transparent environment for contemporary users. This universality and invisibility make it difficult to define what ``the Web'' is today, beyond a ubiquitous infrastructure upon which the critical functions, services, and devices of the digital world operate.

Moreover, the Web, like many digital disciplines, has been overshadowed in research by other technologies that now capture the attention of the computing community, such as machine learning-based artificial intelligence, large-scale data science, and platform-dominated recommendation systems \cite{Gillespie2010}.

\looseness=-1
To these tensions, we must add the contemporary crisis in academic publishing, exacerbated by the arrival of large language models (LLMs) and a hyperproliferation of submissions to conferences and journals. The explosive increase in contributions---many outside the intended scope, hastily produced, or redirected from other venues after negative reviews---has overloaded reviewers and program committees, eroding the quality filters of traditional scientific practice \cite{Ioannidis2023}. The emergence of automated tools for writing, reviewing, and plagiarism detection has introduced new dynamics not yet normatively absorbed, further complicating rigorous evaluation and thematic coherence.

Clearly the area is facing a problem that most researchers are aware of, but whose causal explanation is far from being evident. 
In this poster paper, we present a hypothesis and make an argument for it.\footnote{Qualitative in nature and based on open data sources and the authors' experience in program committees of Web-related venues.} 
\looseness=-1
We claim that the Web, as an object of study, has seen its centrality diluted as a clear research domain, primarily due to its integration into the innumerable subsystems of the digital ecosystem. This fragmentation manifests in the thematic dispersion of submitted work and a loss of coherent disciplinary identity within academic venues that historically placed the Web at their core. The consequence is a mismatch that leads to the acceptance of papers that have no genuine connection to Web-centric research, a phenomenon particularly evident at The Web Conference (formerly WWW) and the International Semantic Web Conference (ISWC).

Thus the aim of this paper is to contribute arguments and perspectives to the necessary open discussion about the Web's place within the current digital landscape. 
Our distinctive contribution is the application of Simon's ``artificial systems'' theory to re-categorize the Web as an environment rather than an instrument. While prior research has viewed the Web as a data source or a platform, our environmental approach helps explain why the discipline is experiencing a ``tragedy of the commons'': the prestige attracts researchers from competitive adjacent fields; this overloads committees and erodes identity; and diminishes the weight of proper Web research.
As illustration, we present the case of the evolution of submissions and topics at The Web Conference, which serves as both a symptom and a symbol of this paradigmatic crisis. 

For researchers of the Web, this crisis is deepened by the advent of AI, a technology that seems to be reconfiguring most traditional digital techniques. The very notions of ``navigation,'' ``interaction,'' or ``information access'' are being reshaped by conversational models and autonomous agents.
This prompts a fundamental question: does the Web remain the privileged medium for human–machine interaction, or is it on its way to becoming a background infrastructure, overshadowed by intelligent interfaces that operate atop it without user awareness?

\looseness=-1
In sum, this article argues that we are witnessing a dual transition: a loss of the Web's academic centrality and a reconfiguration of the scientific system, both driven by the pressures of mass participation, data availability, and ML/LLM platforms. These trends converge in the need to reassess what it means to study ``the Web'' in 2026, what its corresponding disciplinary object is, and what role historical venues such as The Web Conference should play in this new landscape.

This paper is organized into four sections. First, a brief theoretical framework outlining the historical evolution of the Web as a technological object (the ``material'' basis of the crisis). Second, a short analysis of the evolution of Web research topics. Third, an examination of the current situation of the Web Conference. Finally, some reflections and conclusions.

\section{The Technological Landscape: Tools, Machines, and Environments}

To understand the historical and conceptual trajectory of the Web, it is useful to distinguish among different types of technological artifacts. This approach draws inspiration from Herbert Simon's theory of artificial systems \cite{Simon1996}, which emphasizes that artifacts must be understood in relation to their designed purposes and the forms of interaction they support. This foundation allows us to understand technology not merely as tools, machines, and methods, but also as environments, an idea also explored in contemporary theories of technology \cite{Feenberg2017}. Let us briefly review these categories.

\smallskip

\noindent
\textbf{Tools} act as extensions of human agency, expanding existing capabilities without substituting for user action. Their purpose is specific and narrow: a hammer drives nails; a text editor facilitates writing; Photoshop enables image editing. Control is immediate and total, with human intentionality fully governing the operation.
These artifacts (together with machines) form the traditional focus of technological studies, which emphasize functionality, efficiency, and usability. Context is not a central concern. A contribution in this area is evaluated by its performance, a ``property'' of the tool itself.

\smallskip

\noindent
\looseness=-1
\textbf{Machines}, by contrast, introduce partial autonomy and delegation. They automate functions previously requiring human effort or inaugurate capabilities that did not exist before. Following Mumford's classical definition, machines are ``organic systems composed of coordinated parts that perform work through external energy'' \cite{Mumford1967}. This redistribution of agency---part remaining with the user, part delegated to the artifact---complicates human–machine relations, a dynamic especially pertinent to contemporary AI systems. Automobiles, washing machines, and neural-network systems are paradigmatic examples. Their study increasingly intersects with other disciplines, and the context of use becomes relevant precisely because agency is partially delegated. That context becomes a relevant part of their evaluation.

\smallskip

\noindent
\textbf{Environments} constitute a qualitatively distinct category. They are not instruments one wields nor autonomous systems one programs, but spaces designed to be inhabited and within which technological developments occur. They possess their own infrastructures, rules, limitations, and possibilities. Cities, traffic systems, university campuses, and digital platforms are paradigmatic examples. Environments do not do things for the user, nor are they directly controlled. Rather, they define the set of possible actions, shaping behavior through constraints and affordances \cite{Norman2013}. When dealing with environments, the analytical focus necessarily diffuses, as they involve, by design or evolution, multiple stakeholders and manifold disciplines.

\subsection*{The Web as an Environment}

Within this framework, the Web emerges unambiguously as an environment: the universal space in which digital activities occur. This characteristic explains why, just as natural philosophy fractured into specialized disciplines (physics, chemistry, biology), the Web has lost the centrality it held at its inception. Railway engineering was crucial in the nineteenth century, yet today it is a marginal discipline, not due to a lack of importance, but because its object of study has become normalized and integrated into society. The same appears to be happening with the Web.

\looseness=-1
Its environmental nature explains its historical evolution: what was once a central technological object has become a normalized, ubiquitous, and nearly invisible environment. This mirrors the trajectory by which electricity ceased to be an object of fascination and became basic infrastructure \cite{Edwards2003}. Another helpful analogy is road engineering: as railroads and roads became woven into the social fabric, the discipline lost central academic prominence, not due to irrelevance, but to maturity. Studies of digital infrastructure show similar patterns: the most critical systems often disappear from view as they become essential.

Following the classical metaphor of digital highways \cite{Fluckiger1995}, the Web has become the substrate that enables other technologies to flourish. In this sense, the Web is the foundation for tools and machines whose core technologies belong to other disciplines (AI, ML, platforms, privacy), while the Web itself, as an environment, fades into a silent, ubiquitous backdrop.

The question then becomes: How should we evaluate a scientific or technological ``contribution'' to the Web? And who are the proper experts to do so?

\section{The Evolution of Web Research}

The scientific and technological study of the Web has a rich history. We briefly trace its thematic and formal disciplinary evolution as reflected in conferences and journals.

\subsection*{Three Phases of Web Studies}

\paragraph{Phase 1 (1993–2000): Beginnings, Creation and Exploration of the Environment}
Born from HTTP, HTML, and URIs, the so-called ``Web 1.0'' was a network of static linked documents \cite{bernersLee1999}. Users were largely passive consumers of information. Research focused on adapting existing disciplines and technologies to this new space: early browsers, hypertext theory, caching, crawling, and information retrieval. Relevant topics included core protocols, hypertext and navigation, indexing and crawling, and early Web services and machine-to-machine communication. The evaluation of a contribution was clear: how to improve the performance and functionality of the core components that were giving form to this new space.

\paragraph{Phase 2 (2000–2015): Systematic Study and the Social Platform}
The rise of social networks and mobility transformed the Web into a ``read–write'' environment (the so-called ``Web 2.0'' \cite{OReilly2005}). Users became active content producers. Simultaneously, the vision of a machine-readable Web (the Semantic Web) emerged \cite{BernersLee2001}. Research intensified in data mining, social computing, recommender systems, security, and privacy. This was an era of sustained exploration of an already stable infrastructure, developing a characteristic set of tools and methodologies for this dynamic space.

\paragraph{Phase 3 (2015–Present): Subsumption and Fragmentation}
\looseness=-1
The Web has become a stage for broader developments. It gave rise to and was crucial for ``Big Data'' studies, which later acquired a certain independence, reducing the Web to a network of online data sources. Consequently, cutting-edge research no longer concentrates primarily on the Web per se but on phenomena occurring within it, using it as a lens or data source. Alongside the data revolution, other themes have risen to the center, including artificial intelligence (LLMs, graph neural networks), ethics (bias, misinformation), and platform economies. The scale required for research, driven by massive infrastructures and platform monopolies over resources, data, and users, has profoundly shaped the research agenda.

It is particularly in this phase that submissions to Web conferences have grown to large numbers, and the fragmentation of themes has exploded, with a significant influx of contributions from AI and ML disciplines.

\medskip

Finally, it is relevant to note that a new phase may be emerging, marked by the integration of agentic systems and Web-native automation, where the Web becomes both a data substrate and an operational environment for autonomous software agents (a sort of ``Semantic Web++''). This raises a critical question: what should be the touchstone for evaluating contributions in this field from a Web-centric perspective?

\subsection*{The Academic Ecosystem}

These thematic shifts have produced a diverse ecosystem of conferences and journals. Today, Web studies are not concentrated in a single venue but are distributed across an ecosystem reflecting the three strands above:

\paragraph{The Web as a Technological System}
This strand maintains a narrow focus on architecture, protocols, and infrastructure. Key venues include The Web Conference (in its technical tracks), ISWC, and ACM Transactions on the Web (TWEB). These venues often treat the Web as a clearly delimited technical object --one that is frequently assumed as a given rather than problematized.

\paragraph{The Web as a Socio-Technical Ecosystem}
This broader view incorporates human and social layers. Its natural home is ACM Web Science (WebSci), complemented by conferences like SocInfo and ICWSM. Unlike technical venues, the WebSci community was established specifically to reflect upon the Web as a unique, socio-technical phenomenon.

\paragraph{The Web as a Methodological Lens}
Here, the Web is a primary, though not exclusive, example for studying complex and networked information systems. Many conferences in data science and AI now indirectly relate to these Web facets.

\medskip

This specialization signals maturity but also fragmentation: a phenomenon affecting the entire computing discipline. 
We must distinguish, however, the specific case of Web Science (WebSci) \cite{manifesto}, differentiating between unintentional dilution (technical conferences losing focus) and intentional breadth (WebSci’s multidisciplinary nature). 
\looseness=-1
Nevertheless, the central question persists: what exactly constitutes ``the Web'' today? Defining it through technical infrastructure (HTTP, HTML, URI), network topology, or the platforms running over it is no longer sufficient. Today WebSci topics have expanded into a vast multidisciplinary array encompassing digital politics, LLM reliability, cybercrime, and public health. Although this breadth is fundamental to WebSci, the increasing dilution of a clearly defined object remains problematic. In the same way that ``natural philosophy'' fragmented into specialized fields, or ``Physics'' became too broad a designation for a focused conference, the same risk now faces Web studies. While this fuzziness is an inherent challenge for WebSci, in the context of technical conferences, we argue it represents a critical loss of disciplinary identity, as we discuss in the following section.

\section{The Case of the Web Conference Today: Between the Tragedy of the Commons and AI}

\looseness=-1
Returning to the central question: the meaning of the Web as an object of study. What is The Web Conference about today?

Multiple signs point to an ongoing crisis. Researchers increasingly encounter accepted papers that neither require nor meaningfully engage with core Web concepts. It is not rare to find papers that can be fully understood and evaluated without any knowledge of the Web. More strikingly, in the process of reviewing, specialists in Web-centric research often find they have little substantive feedback to offer on these works.

\looseness=-1
Many papers make only marginal contributions to the state of the art (the so-called ``delta-papers''), a phenomenon transversal to many scientific disciplines. The sheer volume of submissions has overwhelmed traditional evaluation processes; for The Web Conference, submissions grew from roughly 1,000 in 2010 to circa 8,000 in 2025. In this context, assessing the genuine relevance of a paper—let alone comparing and balancing disparate contributions—becomes increasingly difficult.

We highlight four interconnected factors that contribute to this situation:

\paragraph{1. The Academic Tragedy of the Commons}
This classic concept serves as a potent metaphor for the conference's current dynamics.

The Attraction of Prestige: As an A* venue (CORE), it attracts researchers from highly competitive adjacent fields like AI and NLP, who seek high-impact publication outlets regardless of their work's relevance to the Web.

Scope Creep: As more AI-centered papers are accepted, the conference's core identity erodes. This creates a negative feedback loop: researchers dedicated to core Web technologies perceive the venue as less relevant and subsequently migrate elsewhere.

Depletion of a Common Resource: The intellectual identity and focus of a conference are a shared resource. Without clear governance and strict scope definition, this resource is depleted by an influx of researchers from other fields, resulting in a venue that is, effectively, ``about everything.''

This crisis is not confined to The Web Conference but is symptomatic of a broader phenomenon: the ``Web'' as an object of study has been so successful that it has been subsumed by the very disciplines it once helped to generate.

\paragraph{2. The Nature of the Web Itself}
\looseness=-1
The above dynamic would be impossible if the Web were not a universal space, an environment capable of containing nearly all digital phenomena. As a modern ``digital nature,'' it is the environment where tools, machines, and methodologies coexist. Each draws on the Web for data, protocols, or techniques, and each contributes to it in some sense. The scope of activities that can be framed as enriching the Web is virtually universal, much like the universal contribution of human activities to the social realm.

\paragraph{3. The Crisis of Data Access} \looseness=-1
A critical material factor driving this conceptual ambiguity is the increasing restriction of access to social media datasets. If datasets are the ``lifeblood'' of our research, the current policy of major platforms to restrict access effectively starves the field. Many researchers previously focused on clear Web-centric science have pivoted to adjacent topics not by choice, but by necessity. Science without data cannot be sustained; consequently, lack of access further accelerates the transition of the Web from a primary object of study to an inaccessible background infrastructure.

\paragraph{4. Artificial Intelligence and the Disruption of All Disciplines}
AI, and particularly LLMs, has disrupted nearly every field of study \cite{Duede2024}, and this one---already in a state of crisis---is no exception. Natural-language interfaces, the statistical refinement of formal systems, and the development of autonomous agents make almost any topic seem tangentially related to the Web. Consequently, a single paragraph tenuously connecting a paper to the Web often suffices to justify twelve remaining pages that are entirely self-contained and disconnected from the essence of the Web as an object of study.

This situation is aggravated by the massive reorientation of funding and industrial priorities toward AI, which leaves Web-centered research comparatively under-resourced and  undervalued.

\section{Reflections and Discussion}

\looseness=-1
This note offers neither a formal proposal nor a ``solution.'' Indeed, what would a ``solution'' mean in this context? Our aim is simply to articulate and elevate an issue that frequently surfaces in conversations among reviewers and attendees of conferences dealing with the Web, and present a
framework to understand the phenomenon.

\looseness=-1
What unites us as Web practitioners is the conviction that the Web is the ``natural'' environment for the digital realm, that is, the foundational space for computation and interaction. Above all, it is the infrastructure that binds the digital world together. Yet, unlike language in society (a universal phenomenon that linguistics has long theorized within robust academic structures), the Web, as a technological artifact of similar universality, remains profoundly under-theorized.\footnote{One of the scarce systematic works in this direction is Halpin \& Monnin \cite{Halpin2014}.}

\looseness=-1
This paper has argued that the Web has completed a full life cycle: from a technological novelty, to a mature environment, and finally, to an object of study dissolving into its own success. The identity crisis of its flagship conference is a direct symptom of this transition. While specialization into focused venues is a natural and healthy sign of maturity, it also presents an opportunity: to forge a renewed sense of purpose around the complex socio-technical problems emerging at the intersection of the Web as a global infrastructure and AI as a transformative force. (One might posit ``AI + Web Infrastructure = Digital Activity'' as the 2025 equivalent of Wirth's ``Algorithms + Data Structures = Programs''.)

From the standpoint of scientific venues, and acknowledging our place within a broader systemic crisis driven by expanded access and AI, we face a critical challenge: How can we preserve meaningful quality filters, remain open to novel ideas from adjacent fields, and still maintain a coherent focus on the Web as our core object of study? In short, how do we sustain a vibrant community dedicated to this foundational environment?

\looseness=-1
The path forward cannot be nostalgic. The future of Web studies lies not in clinging to a narrow definition of the Web’s early novelty, but in deliberately embracing its mature role as the fundamental environment where the major challenges of the digital era are negotiated and resolved. The era of novelty has ended; a period of reflective maturity is beginning. This new era demands fresh conceptual tools to articulate a crucial paradox: what it means to study an environment whose most profound property is its own pervasive invisibility.

 \begin{acks}
We are very grateful for the thoughtful comments and insights from the reviewers.
C. Gutierrez thanks funding from ANID - Millennium Science Initiative Program - Code ICN17\_002, and D. Hernández was funded as part of the project IRIS-HISIT by the State Ministry of Science, Research, and the Arts Baden-Wuerttemberg (grant number: Az.: MWK11-0430.0-1/7/5).
\end{acks}

\enlargethispage{2\baselineskip}
\bibliographystyle{ACM-Reference-Format}
\bibliography{references}

@article{manifesto,
  title={A Manifesto for Web Science @10},
  author={Hall, Wendy and Hendler, Jim and Staab, Steffen},
  journal={Web Science Trust},
  month=dec,
  year={2016},
  url={http://www.webscience.org/manifesto}
}

@article{Ioannidis2023,
  author    = {Ioannidis, John P. A. and Berkwits, Michael and Flanagin, Annette and Bloom, Talya},
  title     = {Peer Review and Scientific Publication at a Crossroads: Call for Research for the 10th International Congress on Peer Review and Scientific Publication},
  journal   = {JAMA},
  volume    = {330},
  number    = {13},
  pages     = {1232--1235},
  year      = {2023},
  doi       = {10.1001/jama.2023.17607},
  url       = {https://doi.org/10.1001/jama.2023.17607}
}

@article{BernersLee2001,
  author    = {Berners-Lee, Tim and Hendler, James and Lassila, Ora},
  title     = {The Semantic Web},
  journal   = {Scientific American},
  volume    = {284},
  number    = {5},
  pages     = {34--43},
  month     = {05},
  year      = {2001},
  publisher = {Scientific American, Inc.}
}

@book{bernersLee1999, 
  author = {Berners-Lee, Tim and Fischetti, Mark},
  title = {Weaving the Web},
  publisher = {HarperCollins},
  year = {1999}
}

@book{Halpin2014, 
  editor    = {Halpin, Harry and Monnin, Alexandre},
  title     = {Philosophical Engineering. Toward a Philosophy of the Web},
  publisher = {Wiley Blackwell},
  year = {2014} 
}

@book{Simon1996, 
  author    = {Simon, Herbert A.},
  title     = {The Sciences of the Artificial},
  edition   = {3},
  publisher = {The MIT Press},
  year = {1996}
}

@incollection{Feenberg2017,
  author    = {Feenberg, Andrew},
  title     = {A Critical Theory of Technology},
  booktitle = {Handbook of Science and Technology Studies},
  editor    = {Felt, Ulrike and Fouché, Rayvon and Miller, Clark A. and Smith-Doerr, Laurel},
  pages     = {635--663},
  publisher = {MIT Press},
  year      = {2017}
}

@book{Mumford1967,
  author    = {Mumford, Lewis},
  title     = {The Myth of the Machine: Technics and Human Development},
  volume    = {1},
  publisher = {Harcourt, Brace \& World},
  address   = {New York},
  year      = {1967}
}

@book{Norman2013,
  author    = {Norman, Donald A.},
  title     = {The Design of Everyday Things: Revised and Expanded Edition},
  edition   = {Revised and Expanded Edition},
  publisher = {Basic Books},
  address   = {New York, NY},
  year      = {2013},
  isbn      = {978-0-465-05065-9}
}

@techreport{Fluckiger1995,
  author      = {Flückiger, François},
  title       = {From World-Wide Web to Information Superhighway},
  institution = {CERN, European Organization for Nuclear Research},
  year        = {1995},
  note        = {Presented at INET'95},
  address     = {Geneva, Switzerland}
}

@incollection{Edwards2003,
  author    = {Edwards, Paul N.},
  title     = {Infrastructure and Modernity: Force, Time, and Social Organization in the History of Sociotechnical Systems},
  booktitle = {Modernity and Technology},
  editor    = {Misa, T. J. and Brey, Ph. and Feenberg, A.},
  publisher = {MIT Press},
  year      = {2003},
  pages     = {185--223}
}

@article{OReilly2005,
  author  = {O'Reilly, Tim},
  title   = {What Is Web 2.0: Design Patterns and Business Models for the Next Generation of Software},
  journal = {O'Reilly Media},
  year    = {2005},
  month   = {09},
  note    = {Originally published on O'Reilly website, September 30, 2005.},
  url     = {http://www.oreilly.com/pub/a/web2.0.html}
}

@article{Duede2024,
  author  = {Duede, Eamon and Dolan, William and Bauer, André and Foster, Ian and Lakhani, Karim},
  title   = {Oil \& Water? Diffusion of AI Within and Across Scientific Fields},
  journal = {arXiv preprint arXiv:2405.15875},
  year    = {2024},
  month   = {05},
  eprint  = {2405.15875},
  archiveprefix = {arXiv},
  primaryclass = {cs.DL}
}

@article{Gillespie2010,
  author  = {Gillespie, Tarleton},
  title   = {The Politics of 'Platforms'},
  journal = {New Media \& Society},
  volume  = {12},
  number  = {3},
  pages   = {347--364},
  year    = {2010},
  doi     = {10.1177/1461444809342178}
}


\end{document}